\documentclass{aastex61}

\submitjournal{ApJ}

\shorttitle{CO multi-line observations of HH~80-81}
\shortauthors{Qiu et al.}

\begin{document}

\title{CO multi-line observations of HH~80-81: a two-component molecular outflow associated with the largest protostellar jet in our Galaxy}

\correspondingauthor{Keping Qiu}
\email{kpqiu@nju.edu.cn}

\author[0000-0002-5093-5088]{Keping Qiu}
\affil{School of Astronomy and Space Science, Nanjing University, 163 Xianlin Avenue, Nanjing 210023, China}
\affil{Key Laboratory of Modern Astronomy and Astrophysics (Nanjing University), Ministry of Education, Nanjing 210023, China}

\author{Friedrich Wyrowski}
\affiliation{Max-Planck-Institut f\"{u}r Radioastronomie, Auf dem H\"{u}gel 69, 53121 Bonn, Germany}

\author{Karl Menten}
\affiliation{Max-Planck-Institut f\"{u}r Radioastronomie, Auf dem H\"{u}gel 69, 53121 Bonn, Germany}

\author{Qizhou Zhang}
\affiliation{Harvard-Smithsonian Center for Astrophysics, 60 Garden Street, Cambridge, MA 02138, U.S.A.}

\author{Rolf G\"{u}sten}
\affiliation{Max-Planck-Institut f\"{u}r Radioastronomie, Auf dem H\"{u}gel 69, 53121 Bonn, Germany}



\begin{abstract}

Stretching a length reaching 10~pc projected in the plane of sky, the radio jet associated with Herbig-Haro objects 80 and 81 (HH~80-81) is known as the largest and best collimated protostellar jet in our Galaxy. The nature of the molecular outflow associated with this extraordinary jet remains an unsolved question which is of great interests to our understanding of the relationship between jets and outflows in high-mass star formation. Here we present Atacama Pathfinder EXperiment CO~(6--5) and (7--6), James Clerk Maxwell Telescope CO~(3--2), Caltech Submillimeter Observatory CO ~(2--1), and Submillimeter Array CO and $^{13}$CO~(2--1) mapping observations of the outflow. We report on the detection of a two-component outflow consisting of a collimated component along the jet path and a wide-angle component with an opening angle of about $30^{\circ}$. The gas velocity structure suggests that each of the two components traces part of a primary wind. From LVG calculations of the CO lines,  the outflowing gas has a temperature around 88~K, indicating that the gas is being heated by shocks. Based on the CO~(6--5) data, the outflow mass is estimated to be a few $M_{\odot}$, which is dominated by the wide-angle component. A comparison between the HH~80--81 outflow and other well shaped massive outflows suggests that the opening angle of massive outflows continues to increase over time. Therefore, the mass loss process in the formation of early-B stars seems to be similar to that in low-mass star formation, except that a jet component would disappear as the central source evolves to an ultracompact H{\scriptsize II} region. 

\end{abstract}


\section{Introduction} \label{sec:intro}

Jets and outflows are found to be ubiquitous in the formation of stars of all the masses \citep[see][for recent reviews]{Frank14,Bally16}. Thanks to their variety of manifestations, e.g., Herbig-Haro (HH) objects,  molecular hydrogen objects (MHOs), radio jets, and molecular outflows, they are observable from X-ray to radio wavelengths. Molecular outflows are often observed in rotational transitions of CO and some other molecules (e.g., SiO). They are of particular interests to our understanding of the earliest stages of star formation, when the central protostars are deeply embedded in gas and dust cores which are invisible at optical to near-infrared wavelengths. These outflows are often the first clear sign of the formation of a new star \citep[e.g.,][]{PhanBao08,Tobin16,Tan16,Feng16}, and provide insights into the mass accretion process as well as the multiplicity of the central protostars \citep[e.g.,][]{Beuther02b,Plunkett15,Hsieh16}.

Outflows in low-mass young stellar objects (YSOs) are far better studied, both observationally and theoretically, compared to their counterparts in the high-mass regime. They have long been thought to be ambient material accelerated by an underlying jet or wind moving at velocities of order 100~km\,s$^{-1}$ \citep[see, e.g.,][and references therein]{Arce07}, but some extremely high velocity structures may originate from the close vicinity of a central protostar \citep[e.g.,][]{Tafalla10,Tafalla15}, and moreover, there is evidence from new observations that molecular outflows could be directly ejected from an accretion disk \citep{Bjerkeli16,Alves17,Lee17,Tabone17,Guedel18}. It has been noted that interaction between a single collimated jet or a wide-angle wind and the ambient cloud could not explain the full range of observed features of molecular outflows \citep{Cabrit97,Lee00,Lee01,Lee02,Arce02}. In particular, high angular resolution observations often show that low-mass protostellar outflows contain a collimated, jet-like component at higher velocities, and a wide-angle, shell-like component at lower velocities \citep{Bachiller95,Gueth99,Palau06,Santiago09,Hirano10,Lee18}. Such two-component outflows could be tracing a laterally stratified primary wind, or an axial jet surrounded by a wide-angle wind, breaking out of a dense infalling envelope \citep{Arce06,Shang06}. The primary jet or wind is launched through the coupling of magnetic fields and dense gas rotation around the central protostar, but the detailed mechanism is not well understood \citep[see][and references therein]{Li14}. Recent Atacama Large Millimeter/submillimeter Array (ALMA) observations suggest that the collimated jet has a launching radius at sub-AU scales on the disk \citep{Lee17}, whereas the wide-angle wind is ejected from a region up to a radial distance of a few tens of AU on the disk \citep{Bjerkeli16,Tabone17}. 

Outflows in high-mass YSOs have sizes and velocity structures similar to those in low-mass outflows, but have orders of magnitude greater masses and energetics \citep{Zhang01,Zhang05,Beuther02a,Bally16}. Based on the statistics of a large sample of CO outflows observed with single-dish telescopes, \citet{Wu04} find that outflows in luminous sources ($>10^3~L_{\odot}$) are systemically less collimated than flows in lower luminosity sources. On the other hand, high angular resolution observations made with millimeter or submillimeter interferometers have detected both highly collimated outflows and wide-angle outflows in high-mass YSOs \citep[e.g.,][]{Shepherd98,Cesaroni99,Qiu09a,Qiu09b,Zhang15}. There is even an explosive, rather than bipolar, outflow in the well-known Orion BN/KL region \citep{Zapata09,Bally17}. Theoretically, it appears to be a consensus in numerical simulations that outflows would be generated during the collapse of a massive cloud core if magnetic fields are included \citep{Banerjee07,Peters11,Hennebelle11,Commercon11}, but the outflow launching zone is not resolved and the simulations were not run long enough to allow a comparison to observations. More recently, a few numerical works focusing on the developing of outflows in high-mass star-forming cores suggest that the disk wind model is applicable to the high-mass regime \citep{Seifried12,Kuiper15,Matsushita18}. Since high-mass YSOs are typically far away from the Sun and tend to reside in crowded clusters, there are few observations capable of constraining the launching of their jets and outflows \citep{Carrasco15,Hirota17}. Many basic properties of outflows in high-mass YSOs, such as the collimation, excitation conditions, evolution, and driving mechanism, are poorly known. The question that whether outflows in high-mass YSOs are scaled up versions of those in low-mass YSOs remains open.

The radio jet associated with HH objects 80 and 81 is driven from a high-mass YSO with a bolometric luminosity of $2\times10^4~L_{\odot}$ at an adopted distance of 1.7~kpc \citep{Rodriguez80,Reipurth88,Marti93}. The jet measured 5.3~pc in projection from HH~80 to a radio source to the north \citep[HH~80~North,][]{Marti93}, and was updated to 7.5~pc with the detection of an outer bow shock beyond HH~80 \citep{Heathcote98} and even larger to 10.3~pc by including a newly detected radio source along the jet path beyond HH~80~North \citep{Masque12}. This makes the HH~80--81 jet far larger than any other YSO jet or HH object known so far.  The jet material moves extremely fast with tangential velocities of $\sim$600--1400~km\,s$^{-1}$ for the inner knots \citep{Marti93,Marti95} and of $\sim$200--400~km\,s$^{-1}$ for the outer knots \citep{Heathcote98,Masque15}. If a proposed inclination angle of $56^{\circ}$ (from the plane of the sky) is taken into account, the jet length and velocity would be further increased by a factor of 1.8 \citep{Heathcote98}. It is also one of the few YSO jets showing non-thermal emissions and is the first detected in linearly polarized synchrotron emission attributed to relativistic electrons \citep{Carrasco10,RK17,Vig18}. The central source of the jet is found to be surrounded by a disk-like structure with a radius of a few 100~AU \citep{Fernandez11b,Girart18}. The \emph{Spitzer} 8~$\mu$m image reveals the wall of a biconical cavity surrounding the radio jet \citep{Qiu08}. All this makes the HH~80--81 radio jet an ideal target for testing whether protostellar jets and outflows in low-mass and high-mass YSOs share a common driving mechanism. However, the nature of the associated outflow is far less clear. Previous single-dish CO low-$J$ observations detected a parsec-sized outflow in the region, but the maps were of low resolutions (16--45$''$) and apparently affected by contaminations from ambient gas, and thus could not resolve the morphology and kinematics of the outflow \citep{Yamashita89,Ridge01,Benedettini04,Wu05}. Existing interferometer CO~(2--1) observations toward the central source of HH~80--81 failed to identify outflow structures associated with the radio jet \citep{Qiu09b,Fernandez13}. Here we present CO multi-line observations covering the central parsec area of the radio jet, aimed at identifying and characterizing the molecular outflow associated with this extraordinary jet. We describe our observations in Section \ref{sec:obs}, and show the results in Section \ref{sec:res}. Discussions on the properties of the HH~80--81 outflow, and its implications on a possible evolutionary picture for massive outflows, are presented in Section \ref{sec:dis}. Finally, a brief summary of this work is given in Section \ref{sec:sum}.

\section{Observations and Data Reduction} \label{sec:obs}
\subsection{APEX Observations} 
We performed CO (6--5) and (7--6) observations on 2010 July 3 with the Atacama Pathfinder EXperiment\footnote{This publication is based on data acquired with the Atacama Pathfinder Experiment (APEX). APEX is a collaboration between the Max-Planck-Institut fur Radioastronomie, the European Southern Observatory, and the Onsala Space Observatory.} (APEX) and its Carbon Heterodyne Array of the MPIfR \citep[CHAMP$^+$,][]{Kasemann06}. CHAMP$^+$ is a dual-color heterodyne array consisting of $2\times7$ pixels for spectroscopy in the 450 and 350~$\mu$m atmospheric windows. Each of the 14 CHAMP$^+$ pixels outputs signals into two 1.5 GHz wide Fast Fourier Transform (FFT) spectrometers configurable for a total bandwidth of 2.4 to 2.8 GHz (corresponding to overlaps of 600 to 200 MHz). We tuned the receiver array to simultaneously observe CO~(6--5) at 691~GHz and CO~(7--6) at 806~GHz, and configured the spectrometers, each divided into 2048 channels,  to have a bandwidth of 2.4 GHz. The APEX beams at these two frequencies are about $9.\!''0$ and $7.\!''7$. We obtained $2'\times1.\!'5$ maps centered at (R.A., Decl.)$_{\rm J2000}$=($18^{\rm h}19^{\rm m}12.\!^{\rm s}1$, $-20^{\circ}47'31''$) with the on-the-fly (OTF) mode. The OTF maps were sampled with $40\times30$ grid cells and a cell size of $3''$, and the long axis was titled by $19^{\circ}$ east of north to follow the orientation of the radio jet. The data were processed with the GILDAS/CLASS package for baseline fitting and subtraction, velocity smoothed into 1~km\,s$^{-1}$ channels, and re-gridded into cell sizes of $4.\!''5$ and $3.\!''85$ (half of the beams) for CO (6--5) and (7--6), respectively. The final data have an intensity scale in $T_{\rm A}^{\ast}$ and the root mean square (RMS) sensitivities are 0.1~K for CO (6--5) and 0.3~K for CO (7--6). For quantitative analyses such as Large Velocity Gradient (LVG) calculations, we convert the intensity scale from $T_{\rm A}^{\ast}$ to the main-beam antenna temperature ($T_{\rm mb}$) with a beam efficiency of 0.41, which was measured toward planets (Jupiter, Mars, and Uranus) in late July 2010.

\subsection{CSO Observations} 
The CO (2--1) observations were undertaken on 2014 July 7 with the Caltech Submillimeter Observatory\footnote{This material is based upon work at the Caltech Submillimeter Observatory, which is operated by the California Institute of Technology.} (CSO) and its 230 GHz receiver. The output signal was processed by a FFT spectrometer which was configured to have a total bandwidth of 1 GHz divided into 8192 channels. The CSO beam at the frequency of CO (2--1) is about $32''$. We made OTF observations to obtain a map with $11\times6$ grid cells and a grid cell size of $16''$. The data were processed with the GILDAS/CLASS package for baseline fitting and subtraction, and velocity smoothed into 1~km\,s$^{-1}$ channels. The calibrated data in $T_{\rm A}^{\ast}$ have an RMS sensitivity of 0.2 K. The intensity scale in $T_{\rm mb}$ could be derived with a beam efficiency of 0.70, following \url{http://www.submm.caltech.edu/cso/receivers/beams.html}.

\subsection{JCMT Observations} 
The CO (3--2) observations were retrieved from the James Clerk Maxwell Telescope\footnote{The James Clerk Maxwell Telescope has historically been operated by the Joint Astronomy Centre on behalf of the Science and Technology Facilities Council of the United Kingdom, the National Research Council of Canada and the Netherlands Organisation for Scientific Research.} (JCMT) archive. The data were taken on 2008 March 25  through the program M08AU19 \citep{Maud15}. A raster map with a size of $7'\times7'$ and a scanning spacing of $7.\!''3$ was obtained with the 16-pixel Heterodyne Array Receiver Program (HARP) and the Auto Correlation Spectral Imaging System (ACSIS), and the latter was configured to have a bandwidth of 1 GHz divided into 2048 channels. The JCMT beam at the frequency of CO (3--2) is about $14.\!''5$. The data were processed with the ORAC-DR pipeline software following the REDUCE\_SCIENCE\_GRADIENT recipe. The calibrated data in $T_{\rm A}^{\ast}$ were velocity smoothed into 1 km\,s$^{^{-1}}$ channels, and the corresponding RMS sensitivity is about 0.2 K. The intensity scale conversion from $T_{\rm A}^{\ast}$ to $T_{\rm mb}$, whenever needed, would use a beam efficiency of 0.64, following \url{http://www.eaobservatory.org/jcmt/instrumentation/heterodyne/harp/}.

\subsection{SMA Observations} \label{subsec:sma}
We carried out Submillimeter Array (SMA)\footnote{The SMA is joint project between the Smithsonian Astrophysical Observatory and the Academia Sinica Institute of Astronomy and Astrophysics and is funded by the Smithsonian Institution and the Academia Sinica.} observations centered at (R.A., Decl.)$_{\rm J2000}$=($18^{\rm h}19^{\rm m}11.\!^{\rm s}0$, $-20^{\circ}48'20''$), approximately the tip of the southwestern lobe of the outflow seen in the APEX CO (6--5) map. The observations were made on 2017 April 12 under excellent weather conditions with the atmospheric opacity at 225~GHz ranging from 0.06 to 0.08. The array was in the Compact configuration with 7 antennas available during the observations. Each SMA antenna is now equipped with four receivers, namely 230~GHz, 240~GHz, 345~GHz, and 400~GHz receivers, and allows dual-receiver operations. We used 230~GHz and 240~GHz receivers, and both receivers were tuned to the same frequency coverage, $\sim$213.5--221.5 GHz in the lower sideband and $\sim$229.5--237.5 GHz in the upper sideband, to improve the signal-to-noise ratios for spectral line observations. The frequency setup covered CO~(2--1) and $^{13}$CO~(2--1). The newly commissioned SWARM (SMA Wideband Astronomical ROACH2 Machine) correlator was used to provide a uniform spectral resolution of 140 kHz across 8 GHz per sideband per receiver. We smoothed the data by a factor of 4, resulting in a 560 kHz resolution, corresponding to $\sim$0.73~km\,s$^{-1}$ at 230 GHz. 3C279 and Callisto were observed as the bandpass and flux calibrators, respectively. The time-dependent gain variations were monitored through interleaving observations of two quasars, J1733-130 and J1924-292. We calibrated the data with the IDL MIR package\footnote{\url{https://github.com/qi-molecules/sma-mir}}, and then output the calibrated visibilities to MIRIAD for imaging. The final CO~(2--1) map has a synthesized beam with a full-width-half-maximum (FWHM) size of $4.\!''9\times2.\!''5$ and a position angle (PA) of $-23^{\circ}$, and the $^{13}$CO~(2--1) map has a synthesized beam of $5.\!''3\times2.\!''3$ with a PA of $-26^{\circ}$. The RMS sensitivity is about 0.14~Jy\,beam$^{-1}$ (or 0.27~K) at a velocity resolution of 0.73~km\,s$^{-1}$.

We summarize in Table \ref{tab:lines} the key parameters of each observed CO line, including the frequency, equivalent temperature of the upper level energy, angular resolution, velocity resolution, and RMS sensitivity.

\begin{deluxetable*}{crrcccc}
\tablecaption{Key parameters of the observed CO lines \label{tab:lines}}
\tablecolumns{7}
\tablewidth{0pt}
\tablehead{
\colhead{Transition} & \colhead{Frequency} & \colhead{$E_{\rm up}$/k} & \colhead{Telescope} & \colhead{Angular Resolution} & \colhead{Velocity Resoluion} & \colhead{RMS Sensitivity\tablenotemark{a}} \\
\colhead{} & \colhead{(GHz)} & \colhead{(K)} & \colhead{} & \colhead{(arcsec)} & \colhead{(km\,s$^{-1}$)} & \colhead{(K)}
}
\startdata
$J=2$--1 & 230.538 &   16.6 &   CSO & 32\arcsec & 1.0 & 0.2 \\
$J=2$--1 & 230.538 &   16.6 &   SMA & 4.\arcsec9$\times$2.\arcsec5 & 0.73 & 0.27 \\
$J=3$--2 & 345.796 &   33.2 & JCMT & 14.\arcsec5 & 1.0 & 0.2 \\
$J=6$--5 & 691.473 & 116.2 & APEX & 9.\arcsec0 & 1.0 & 0.1 \\
$J=7$--6 & 806.652 & 154.9 & APEX & 7.\arcsec7 & 1.0 & 0.3 \\
\enddata
\tablenotetext{a}{Measured in $T_{\rm A}^{\ast}$, except that for the SMA, measured in the brightness temperature.}
\end{deluxetable*}

\section{Results} \label{sec:res}
\subsection{Single-dish CO multi-transition observations}

\begin{figure}
\plotone{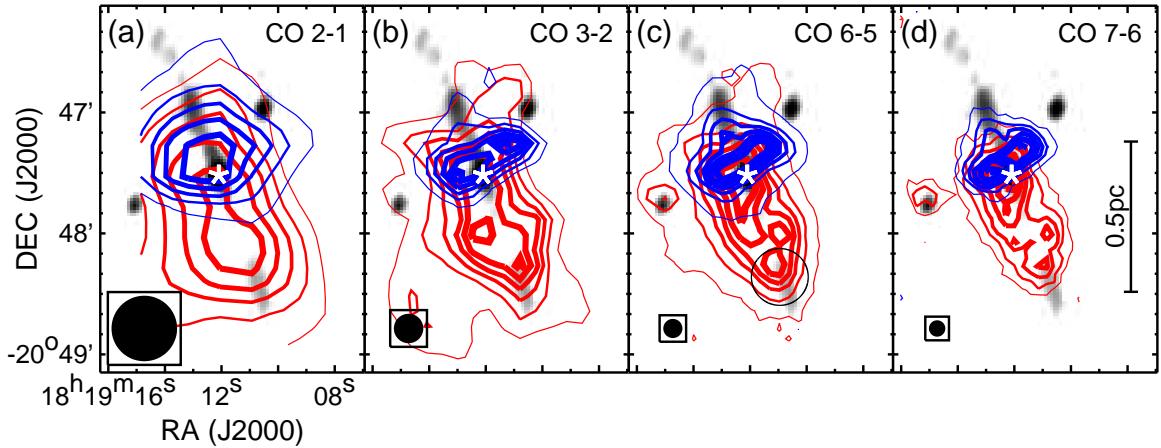}
\caption{Velocity integrated emissions in CO lines, shown in blue and red contours for the blueshifted emission of 3 to 9~km\,s$^{-1}$ and redshifted emission of 15 to 18~km\,s$^{-1}$, respectively. (a) The CSO CO (2--1) map, contouring in steps of 15\% from 30\% to 90\% of the peaks of 36.0 and 32.9~K\,km\,s$^{-1}$ for the blue and red lobes, respectively. (b) The JCMT CO (3--2) map, contouring in steps of 15\% from 20\% to 95\% of the peaks of 64.0 and 42.7~K\,km\,s$^{-1}$ for the blue and red lobes, respectively. (c) The APEX CO (6--5) map, contouring in steps of 15\% from 10\% to 100\% of the peaks of 52.9 and 33.5~K\,km\,s$^{-1}$ for the blue and red lobes, respectively; a circle outlines the area for computing the parameters of the collimated component of the redshifted lobe (see Section \ref{subsec:mass} for details). (d) The APEX CO (7--6) map, contouring in steps of 15\% from 20\% to 95\% of the peaks of 40.0 and 23.7~K\,km\,s$^{-1}$ for the blue and red lobes, respectively. Note that the contour thickness is proportional to the contour levels. In each penal, the grayscale shows the VLA 6~cm image of the radio jet; hereafter, an asterisk denotes the central source of the radio jet; a filled circle in the lower left corner indicates the beam size accordingly. \label{fig:int_maps}}
\end{figure}

Our single-dish observations made with the CSO, JCMT, and APEX cover the inner $\sim$1~pc of the HH~80-81 radio jet.  Figure \ref{fig:int_maps} shows maps of velocity integrated emissions in CO~(2--1), (3--2), (6--5), and (7--6), with angular resolutions of $32''$, $14.\!''5$, $9''$, and $7.\!''7$, respectively. A northeast-southwest (NE-SW) outflow with a projected length of about 0.8~pc and a P.A. of about $19^{\circ}$ is detected in all the maps. Compared to the Very Large Array (VLA) 6~cm observations, the outflow is clearly associated with the radio jet. The outflow appears increasingly collimated in maps from low- to mid-$J$ transitions and from low to moderately high angular resolutions. To examine whether the variation in the outflow morphology is purely due to the resolution effect, we convolve the CO (6--5) and (7--6) maps to the resolution of the CO (3--2) map, compare the maps in three lines, and find that the outflow does appear more collimated in higher excitation lines (see Appendix~\ref{append:convolved}). The outflow is bipolar, but very asymmetric, having a $\sim$0.5~pc lobe in the SW and a much shorter, stub-like structure in the NE. This is likely due to an inhomogeneous density structure of the cloud gas around the central source. Another noticeable characteristic of the outflow is that the emission is only detected at relatively low velocities, with $-9~{\lesssim}~v~{\lesssim}~6$~km\,s$^{-1}$, where $v$ is the outflow velocity with respect to the cloud systemic velocity of 11.8~km\,s$^{-1}$ \citep{Fernandez11b}. In the CO~(3--2), (6--5), and (7--6) maps, the blueshifted emission is dominated by an elongated structure in a northwest-southeast orientation, which is mostly attributed to other outflows unrelated to the HH~80-81 radio jet \citep{Qiu09b,Fernandez13}, and will not be further discussed in this work. 

\begin{figure}[htb!]
\epsscale{0.5}
\plotone{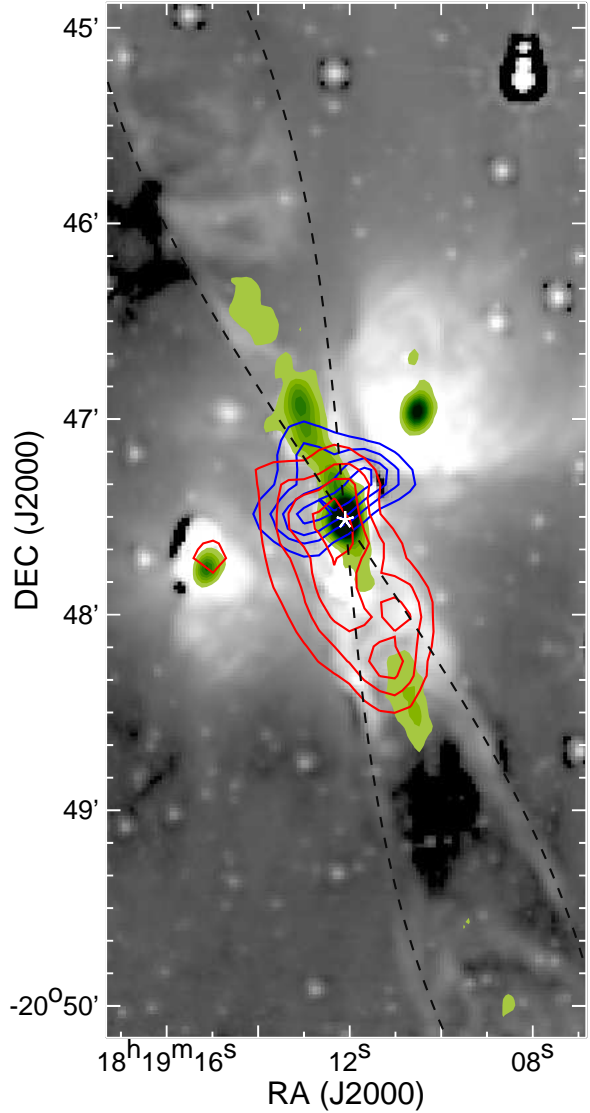}
\caption{An overview of the {\it Spitzer} IRAC, APEX, and VLA observations. The grayscale background shows the IRAC 8~$\mu$m image; we apply a high-pass filtering to the image to highlight the wide-angle wall, which is outlined with two dashed lines intersecting at the central source. Blue and red contours show the APEX CO~(6--5) map, which is the same as Figure~\ref{fig:int_maps}(c), but with contour levels starting from 30\% and continuing in steps of 20\% of the peak emission. Filled green contours show the VLA 6~cm continuum map starting and continuing in steps of 60~$\mu$Jy\,beam$^{-1}$. \label{fig:overview}}
\end{figure}

\begin{figure}
\epsscale{1}
\plotone{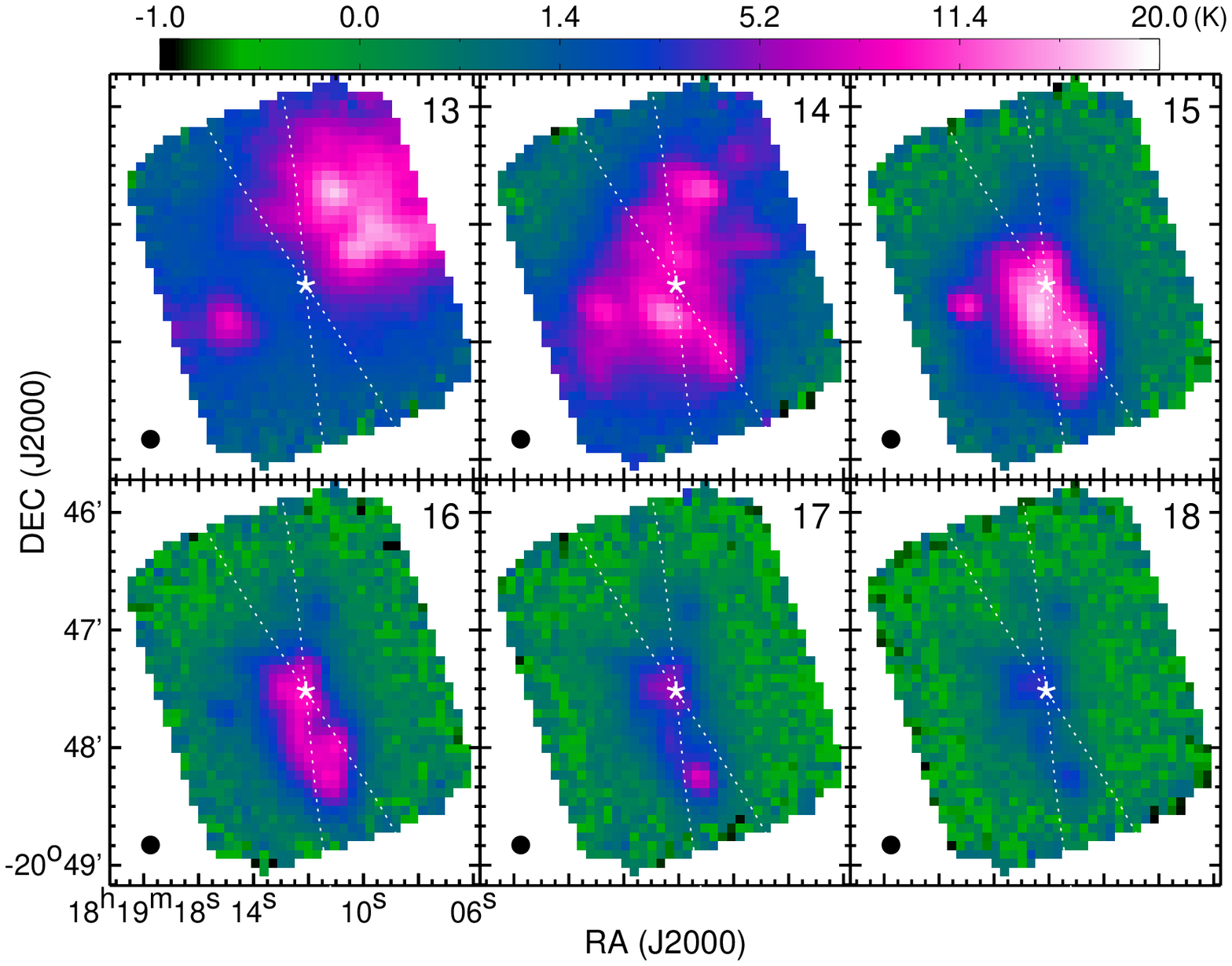}
\caption{Velocity channel maps of the CO~(6--5) emission at 13 to 18~km\,s$^{-1}$. The pseudo color scale, as indicated by a color bar on top of the figure, visualizes intensities from $-1.0$ to 20.0~K in $T_{\rm A}^{\ast}$. Other symbols are the same as those in Figure~\ref{fig:overview}. \label{fig:CO65_chan}}
\end{figure}

Focusing on the JCMT and APEX maps of the SW lobe, the molecular outflow shows a conical, wide-angle structure within a distance of $\sim$0.25~pc from the central source, and appears to re-collimate further out with the tip lying on the axis of the radio jet. To quantify the opening angle of a wide-angle structure around the radio jet, we revisit the {\it Spitzer} IRAC observations \citep{Qiu08}, and measure an opening angle of $\sim$28$^{\circ}$ \footnote{Detailed discussions on various emission mechanisms for an outflow seen in the IRAC bands, as well as a description of the IRAC observations of the HH~80--81 outflow, are presented in \citet{Qiu08}.}. A comparison between the mid-IR cavity, the radio jet, and the CO outflow is shown in Figure \ref{fig:overview}. It seems that the molecular outflow is associated with both the highly collimated jet and the wide-angle cavity wall. Figure \ref{fig:CO65_chan} shows the velocity channel maps of the CO~(6--5) emission from 13 to 18~km\,s$^{-1}$. In channels of 14--18~km\,s$^{-1}$, the emission within a distance of $\sim$0.25~pc from the central source traces a wide-angle component with an opening angle roughly consistent with that of the cavity wall seen in the IRAC 8~$\mu$m image. Meanwhile, the tip of the SW lobe is seen as a clump at a distance of $\sim$0.5~pc from the central source in channels of 16--18~km\,s$^{-1}$; the clump lies on the radio jet axis, suggesting the presence of a centrally collimated component in the molecular outflow. The channel maps of the CO~(3--2) and (7--6) emissions show similar results (see Appendix~\ref{append:chan}).

\subsection{SMA CO and $^{13}$CO (2--1) observations}
\begin{figure}
\epsscale{1}
\plotone{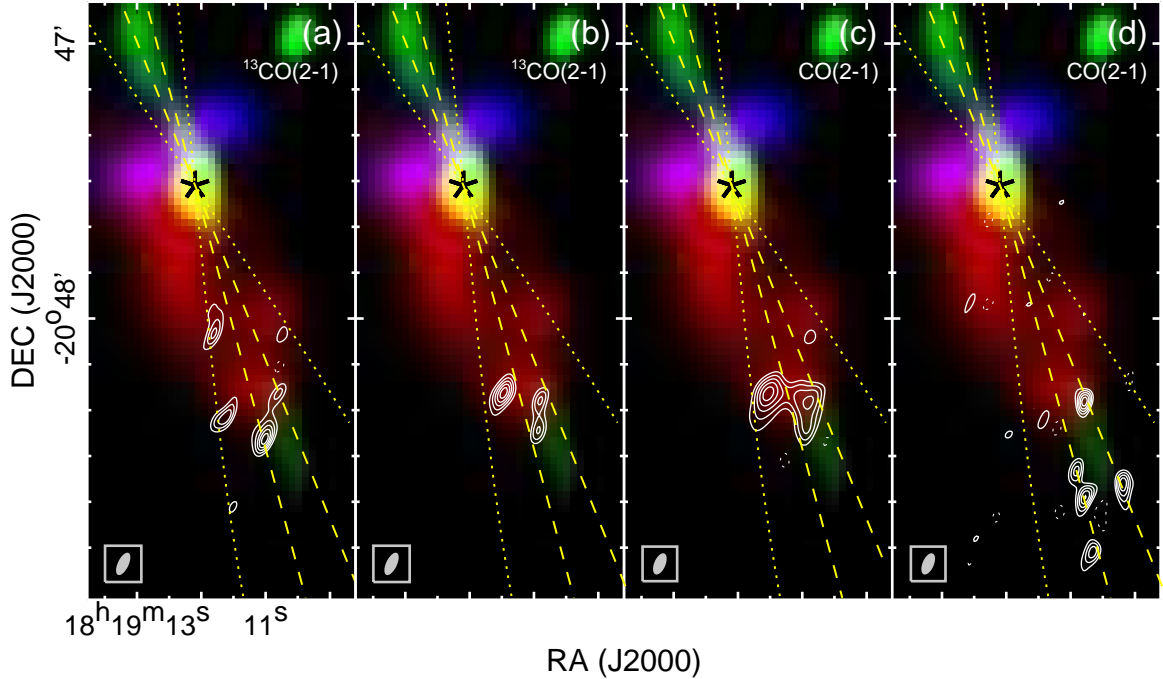}
\caption{Contour maps of $^{13}$CO~(2--1) and CO~(2--1) emissions observed with the SMA. (a) $^{13}$CO~(2--1) emission integrated from 15.38 to 16.11~km\,s$^{-1}$, contouring from 30\% to 90\%, by 15\%, of the peak of 3.19~Jy\,beam$^{-1}$\,km\,s$^{-1}$. (b) $^{13}$CO~(2--1) emission integrated from 16.84 to 18.30~km\,s$^{-1}$, contouring from 30\% to 90\%, by 15\%, of the peak of 3.22~Jy\,beam$^{-1}$\,km\,s$^{-1}$. (c) CO~(2--1) emission integrated from 14.66 to 18.30~km\,s$^{-1}$, contouring from 30\% to 90\%, by 15\%, of the peak of 40.62~Jy\,beam$^{-1}$\,km\,s$^{-1}$. (d) CO~(2--1) emission integrated from 19.02 to 19.75~km\,s$^{-1}$, contouring from 30\% to 90\%, by 15\%, of the peak of 1.87~Jy\,beam$^{-1}$\,km\,s$^{-1}$. In each panel, the background image shows the CO~(6--5) outflow and the radio jet the same as those shown in Figure~\ref{fig:overview}, but with the red, blue lobes of the CO~(6--5) outflow, and the radio jet coded in red, blue, and green, respectively; two dotted lines outline the outflow cavity wall seen in the IRAC 8~$\mu$m image, and two dashed lines depict the range of PAs of the radio knots newly detected in \citet{RK17}; a filled ellipse in the lower left corner shows the synthesized beam at FWHM. \label{fig:int_sma}}
\end{figure}

Previous interferometric observations toward the central source failed to unveil the outflow associated with the radio jet. This is not surprising now as we know that the outflow velocity is not very high and the low velocity CO emission around the central source is dominated by complicated structures composed of multiple outflows and ambient cloud gas. Guided by the outflow maps shown in Figure \ref{fig:int_maps}, we performed new SMA observations toward the tip of the SW lobe. Figure \ref{fig:int_sma} shows contour maps of the velocity integrated emissions in $^{13}$CO and CO~(2--1), along with a color-composite image of the CO~(6--5) outflow and the radio jet. The cavity wall seen in the {\it Spitzer} IRAC image is delineated by two dotted lines in Figure \ref{fig:int_sma}. Most recently, new sensitive and high angular resolution observations resolve the emission knots of the HH~80-81 radio jet into multiple components \citep{RK17}. Of our particular interests are the radio knots around the SW tip of the CO~(6--5) outflow, which have PAs within a range outlined by two dashed lines in Figure \ref{fig:int_sma} \citep[also see Figure 2 in][]{RK17}. In Figure \ref{fig:int_sma}(a), the $^{13}$CO~(2--1) emission at $v=3.58$--4.31~km\,s$^{-1}$ reveals molecular structures along both the cavity wall and the radio jet. In Figures \ref{fig:int_sma}(b) and \ref{fig:int_sma}(c), the $^{13}$CO~(2--1) emission at $v=5.04$--6.50~km\,s$^{-1}$  and the CO~(2--1) emission at $v=2.86$--6.50~km\,s$^{-1}$ trace molecular gas along and closely around the radio jet. Figure \ref{fig:int_sma}(d) shows the CO~(2--1) emission at $v=7.22$--7.95~km\,s$^{-1}$, which reveals molecular structures all along the radio jet at larger distances from the central source ($\gtrsim0.5$~pc). The SMA arc~second resolution observations confirm the presence of molecular outflow gas associated with both the wide-angle cavity wall and the collimated jet. And in particular, the highest velocity ($v\sim7$--8~km\,s$^{-1}$) CO emission is clearly associated with the radio jet. 

\begin{figure}[ht!]
\epsscale{.8}
\plotone{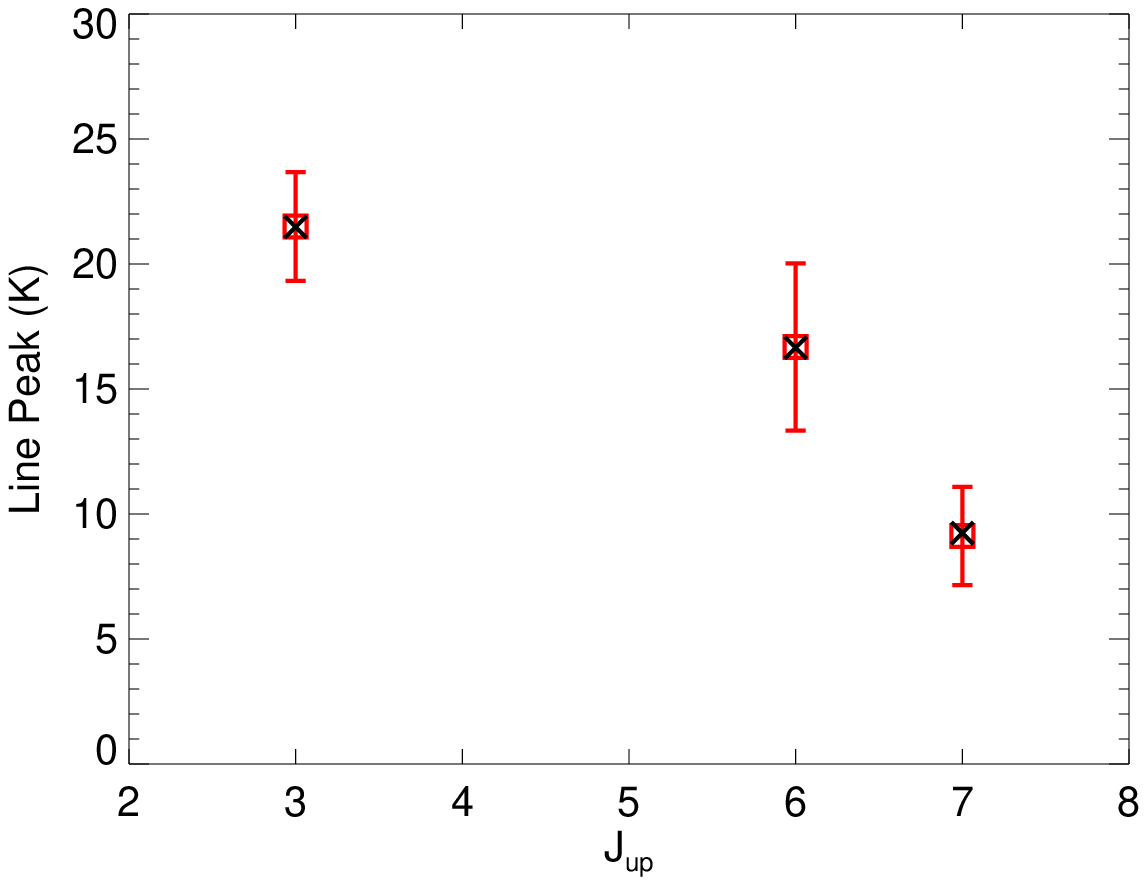}
\caption{CO~(3--2), (6--5), (7--6) line peaks (J$_{\rm up}=3,6,7$) toward the tip of the SW lobe of the outflow, with the observations shown in square symbols along with error bars taking into account both the flux calibration uncertainties and the RMS noise levels, and the best-fit LVG model shown in cross symbols. \label{fig:lsed}}
\end{figure}

\subsection{Large Velocity Gradient calculations of the outflow temperature and density}
We investigate the excitation conditions of the outflow gas by performing LVG calculations of the CO line spectral energy distribution (LSED) using the RADEX code \citep{vdTak07}. To avoid contamination from the ambient cloud gas and other outflows from nearby high-mass protostars, we measure the CO line peaks toward the tip of the SW lobe. The H$_2$ temperature ($T_{\rm kin}$), density ($n_{\rm H_2}$), and ratio of the CO column density to the line width ($N_{\rm CO}/{\Delta}V$), are derived by fitting the observed line peaks with the LVG models with a $\chi^2$-minimization grid search algorithm, where $\chi^2=\sum(T_{\rm R}-T_{\rm mb})^2/\sigma^2$, $T_{\rm R}$ is the modeled line peak, $T_{\rm mb}$ is the measured line peak, and $\sigma$ is the RMS level accordingly. Considering that the CO~(2--1) map has a beam size more than two times larger than the other three lines and to mitigate the beam dilution effect, we fit the CO~(3--2), (6--5), and (7--6) lines after convolving the CO~(6--5), (7--6) maps to the CO~(3--2) beam. From Figure~\ref{fig:lsed}, the best-fit LVG model matches the observations very well, and yields $T_{\rm kin}=88$~K, $n_{\rm H_2}=3.3\times10^4$~cm$^{-3}$, and $N_{\rm CO}/{\Delta}V=1.3\times10^{16}$~~cm$^{-2}$~(km\,s$^{-1}$)$^{-1}$. The uncertainty in the measured line peaks is dominated by the absolute flux calibration errors. If we conservatively adopt a flux calibration uncertainty of 10\% for CO~(3--2) and 20\% for CO~(6--5) and (7--6), the best-fit LVG models indicate that $T_{\rm kin}\sim57$--112~K, $n_{\rm H_2}\sim(3.3$--$7.8)\times10^4$~cm$^{-3}$, and $N_{\rm CO}/{\Delta}V\sim(1.2$--$1.6)\times10^{16}$~~cm$^{-2}$~(km\,s$^{-1}$)$^{-1}$. Thus the outflow gas is much warmer than the ambient quiescent gas. Being about 0.5~pc from the central high-mass protostar, the gas is presumably heated by shocks which are created as the fast jet or wind impinges on the ambient cloud. 

\subsection{The outflow mass and energetics} \label{subsec:mass}
We calculate the mass of the warm outflow gas with the CO~(6--5) data. Provided that the blueshifted emission is dominated by other outflows unrelated to the radio jet, we calculate the mass from the redshifted emission only. By assuming optically thin emission and adopting a canonical CO-to-H$_2$ abundance ratio of $10^{-4}$, we obtain $$M_{\rm red}(v)=5.88\times10^{-9}\,{\rm exp}(116.13/T_{\rm ex})\,(T_{\rm ex}+0.92)\,d_{\rm kpc}^2\,{\Delta}v\,{\int}T_{\rm mb}\,ds$$ where $M_{\rm red}(v)$ is the mass in $M_{\odot}$ of the redshifted outflow within a velocity interval of $v~{\rightarrow}~v+{\Delta}v$, $T_{\rm ex}$ is the excitation temperature, $d_{\rm kpc}$ is the source distance in kpc, and $T_{\rm mb}$ is the measured main-beam antenna temperature and is integrated over an area (measured in arc~second$^2$) encompassing the outflow. We adopt $T_{\rm ex}=88$~K based on the above LVG calculations, and obtain a mass of 1.6~$M_{\odot}$ for the gas at 14--18~km\,s$^{-1}$. In case that the CO~(6--5) emission has a moderate optical depth, $\tau$, the mass estimate should be corrected by a factor of $\tau/(1-e^{\tau})$. The above LVG calculations give an optical depth of 0.68 for the CO~(6--5) emission toward the tip of the SW lobe, and if we take it as an average optical depth for the entire redshifted/SW lobe, the mass of the outflow at 14--18~km\,s$^{-1}$ amounts to 2.2~$M_{\odot}$. The outflow momentum and energy are 6.8~$M_{\odot}$\,km\,s$^{-1}$ and $2.4\times10^{44}$~erg, respectively. The outflow dynamical timescale, derived from the outflow length ($\sim$0.5~pc) divided by the maximum outflow velocity ($\sim$7~km\,s$^{-1}$), is about $7\times10^4$~yr, which is consistent with that of the accretion phase of the central protostar, $(7$--$11)\times10^4$~yr, and compatible with that of the dynamical age of the radio jet, $>9\times10^3$~yr \citep{Masque12}.  Consequently, the mass loss rate is $\sim3\times10^{-5}~M_{\odot}$\,yr$^{-1}$ and the outflow mechanical force is about $\sim1\times10^{-4}~M_{\odot}$~km\,s$^{-1}$\,yr$^{-1}$. All the calculated parameters are listed in Table \ref{tab:outflow}. The emission taken into account is confined within a polygon encompassing the outflow seen in Figure~\ref{fig:int_maps}, and from Figure~\ref{fig:CO65_chan}, the contamination from other outflows around the central source should be minor. Since the outflow is very asymmetric (Figure~\ref{fig:int_maps}), the mass of the blueshifted/NE lobe should be small compared to that of the redshifted/SW lobe. Thus, we expect that our estimates of the outflow mass and energetics represent the lower limits of the parameters.

Given that the outflow shows a two-component structure, it is of interests to evaluate which component, collimated or wide-angle, is dominating the outflow mass and energetics. The emission shown in Figure \ref{fig:int_sma}(d) is coming from the central collimated component, but the SMA observations were made toward the tip of the outflow lobe, and did not cover the entire lobe (Section \ref{subsec:sma}). Moreover, the clumpy structures seen in Figures \ref{fig:int_sma} indicate that the images are affected by the spatial filtering effect of the interferometer. Thus it is difficult to obtain a reasonable estimate of the gas mass of the collimated component with the SMA data. Alternatively, by carefully examining the APEX CO (6--5) and (7--6) channel maps, we find that the two components in the redshifted lobe could be roughly separated, and in particular, the collimated component is dominated by a distant clump at 15 to 18 km\,s$^{-1}$. We thus estimate the mass of the collimated component based on the CO (6--5) data. We make the calculations for the emission at 15--18~km\,s$^{-1}$ in a circular area as outlined in Figure \ref{fig:int_maps}(c), adopting the same excitation temperature and optical depth as the above, and derive a mass of 0.2~$M_{\odot}$. The other parameters are also computed (see Table \ref{tab:outflow}). It is clear that the mass and energetics of the central collimated component are only about 10\% of those of the entire lobe, indicating that the wide-angle component of the outflow is dominating the mass loss and momentum ejection to the ambient cloud.

\begin{deluxetable*}{ccccccccc}
\tablecaption{Calculated parameters of the redshifted lobe and its central collimated component \label{tab:outflow}}
\tablecolumns{9}
\tablewidth{0pt}
\tablehead{
\colhead{Component} & \colhead{Mass} & \colhead{Moment} & \colhead{Energy} & \colhead{Length} & \colhead{Velocity} & \colhead{Time scale} & \colhead{Mass loss rate} & \colhead{Mechanical force} \\
\colhead{} & \colhead{($M_{\odot}$)} & \colhead{($M_{\odot}$\,km\,s$^{-1}$)} & \colhead{(erg)} & \colhead{pc} & \colhead{(km\,s$^{-1}$)} & \colhead{(yr)} & \colhead{($M_{\odot}$\,yr$^{-1}$)} & \colhead{($M_{\odot}$~km\,s$^{-1}$\,yr$^{-1}$)}
}
\startdata
Redshifted lobe & 2.2 & 6.8 & $2.4\times10^{44}$ & 0.5 & 7 & $7\times10^4$ & $3\times10^{-5}$ & $1\times10^{-4}$ \\
Collimated & 0.2 & 0.8 & $3.3\times10^{43}$ & 0.5 & 7 & $7\times10^4$ & $3\times10^{-6}$ & $1\times10^{-5}$ \\
\enddata
\tablecomments{The outflow mass and energetics may represent the lower limits of the parameters (see the discussion in Section \ref{subsec:properties} for more details).}
\end{deluxetable*}

\section{Discussion} \label{sec:dis}

\subsection{Morphology, mass, and energetics of the outflow} \label{subsec:properties}
The HH~80--81 radio jet stands out as the largest and most powerful YSO jet in our Galaxy. The molecular outflow has also been mapped by several groups using single-dish CO~(1--0) and (2--1) observations \citep{Yamashita89,Ridge01,Benedettini04,Wu05}\footnote{\citet{Maud15} estimated the outflow mass and energetics based on the JCMT CO~(3--2) data, but did not provide a map of the outflow.}. These observations do not have sufficient angular resolutions to resolve the outflow morphology, or to distinguish the HH~80--81 outflow from other outflows around the central source. In addition, since the outflow has relatively low velocities ($<10$~km\,s$^{-1}$), low-$J$ CO maps could be easily contaminated by the ambient cloud. This is also the main reason that existing SMA observations centered at the central source failed to disentangle the outflow structure associated with the radio jet; the SMA CO~(2--1) maps at low velocities suffered from side lobes and missing flux due to inadequate $(u,v)$ coverage \citep{Qiu09b,Fernandez13}. Based on the Nobeyama 45~m telescope CO~(1--0) map at a resolution of $16''$ (the highest resolution reached by previous observations of the outflow), the outflow has been thought to have a wide opening angle of $40^{\circ}$ and dose not re-collimate \citep{Yamashita89,Shepherd05,Arce07}, though the outflow axis is misaligned by about $30^{\circ}$ from the radio jet axis. A detailed comparison between the Nobeyama CO~(1--0) map \citep[Figure 2 in][]{Yamashita89} and the JCMT CO~(2--1) and (3--2) maps \citep[Figure 2 in][and Figure \ref{fig:int_maps}b in this work]{Ridge01} indicates that the redshifted emission in the Nobeyama map is contaminated by a minor structure to the southeast seen in the JCMT maps, which along with the true redshifted lobe of the HH~80--81 outflow mimics a wide angle structure. The blueshifted emission of the Nobeyama map is presumably contaminated by other outflows around the central source \citep[see][]{Qiu09b,Fernandez13}. Our APEX maps have a higher resolution than those of previous single-dish observations, and the relatively high excitation conditions of the CO~(6--5) and (7--6) lines help to mitigate contamination from ambient quiescent clouds. Thus the APEX maps unambiguously reveal an outflow associated with the radio jet, and the outflow is overall moderately collimated, and does re-collimate at a distance of $\sim$0.5~pc from the central source. The CO~(6--5) velocity channel maps show that the outflow can be decomposed into two components: a wide-angle component immediately encompassing the cavity wall seen in the \emph{Spitzer} 8~$\mu$m image within $\sim$0.25~pc from the central source, and a collimated component seen as a tip lying about 0.5~pc from the central source on the radio jet axis. The latter is the first detection of the molecular counterpart of the radio jet. The ``two-component'' nature of the outflow is further confirmed by the SMA CO and $^{13}$CO (2--1) maps, which reveal molecular knots and clumps both along the precessing radio jet and along the cavity wall. 

The outflow mass derived from the CO~(6--5) data is about $2.2~M_{\odot}$ for the redshifted lobe at 14--18~km\,s$^{-1}$, and is dominated by the wide-angle component. This is an estimate of the lower limit considering that the blueshifted lobe is not taken into account. The outflow mass ranges from  27 to 570~$M_{\odot}$ in previous single-dish low-$J$ CO observations \citep{Yamashita89,Ridge01,Benedettini04,Wu05,Maud15}, which is one to two orders of magnitude greater than our estimate. The discrepancy could be attributed to several reasons: previous estimates adopted a lower $T_{\rm ex}$ (12--40~K) and greater $\tau$ ($>1$) for the CO emissions; previous low resolution CO maps were apparently contaminated by the ambient cloud and other outflows in the region; our CO~(6--5) map probes a warmer and inner part of the outflow. Thus, whereas the outflow mass could be significantly overestimated in previous studies based on low resolution and low-$J$ CO observations, our correction of the optical depth effect with $\tau=0.68$ may underestimate the outflow mass by a factor of a few. Also considering the uncounted contribution from the blueshifted lobe (though it is minor based on Figures~\ref{fig:int_maps}--\ref{fig:CO65_chan}), we expect that the total mass of the CO~(6--5) outflow to be on the order of a few to 10~$M_{\odot}$. Consequently, the mass loss rate reaches $10^{-4}~M_{\odot}$\,yr$^{-1}$, the outflow mechanical force falls in the range of $10^{-4}$ to $10^{-3}~M_{\odot}$~km\,s$^{-1}$\,yr$^{-1}$, and the outflow energy amounts to $10^{45}$ erg. Considering empirical correlations between outflow parameters and YSO luminosities derived from low-$J$ CO surveys of molecular outflows \citep{Beuther02a,Wu04,Zhang05,Maud15}, the time-averaged parameters (the mass loss rate and the mechanical force) of the HH~80--81 outflow appear to be a bit low, but still within the uncertainties, for a $10^4~L_{\odot}$ source.  In this sense, the HH~80--81 outflow is consistent with a scaled up version of low-mass protostellar outflows \citep{Bachiller95,Gueth99,Palau06,Santiago09,Hirano10,Lee18}, showing a similar morphology but higher mass and energetics.

\subsection{On the relationship between the radio jet and the molecular outflow}
\begin{figure}[ht!]
\epsscale{.8}
\plotone{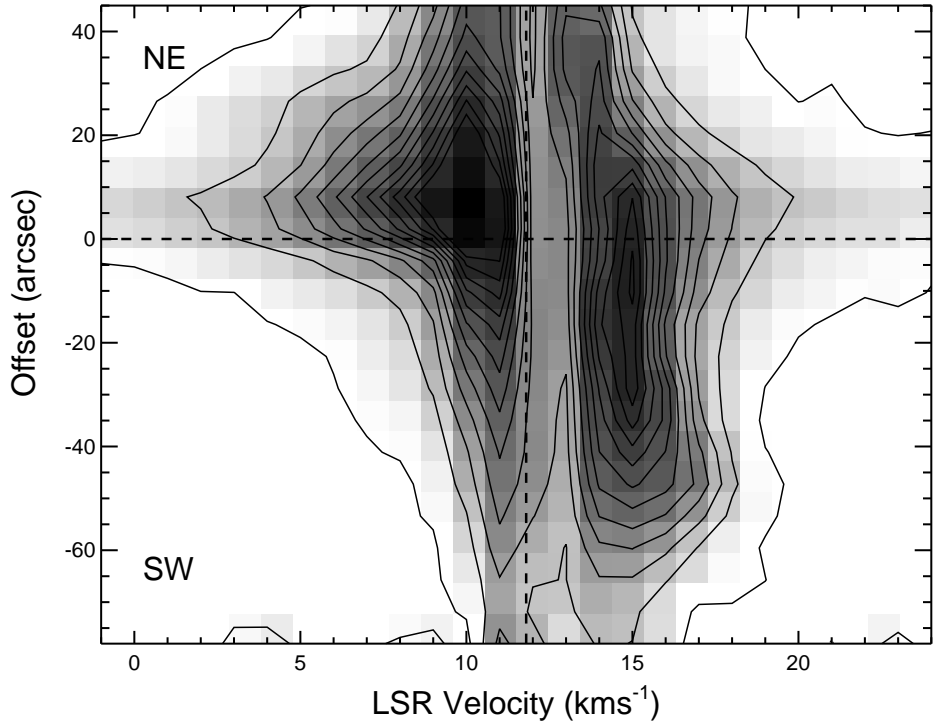}
\caption{APEX CO~(6--5) PV diagram cut along the radio jet axis. Contour levels start at 0.4~K and increase in steps of 2~K, and the grayscale image stretches from 0.4 to 19.6~K in a logarithmic scale. A horizontal dashed line indicates the position of the central source, and a vertical dashed line marks the cloud velocity of 11.8~km\,s$^{-1}$. \label{fig:pv}}
\end{figure}

The HH~80--81 radio jet has been well studied over the past decades. Could the outflow be driven by the fast jet? The mass flux in the ionized jet is estimated to be $(0.6$--$1)\times10^{-6}~M_{\odot}$\,yr$^{-1}$, leading to a momentum rate of $(0.6$--$1)\times10^{-2}~M_{\odot}$\,km\,s$^{-1}$\,yr$^{-1}$ by adopting a jet ejection velocity of $\sim$1000~km\,s$^{-1}$ and an ionization fraction of 0.1 \citep{Carrasco12,RK17}. Thus the thrust available from the fast jet is at least an order of magnitude greater than the mechanical force of the molecular outflow ($10^{-4}$--$10^{-3}~M_{\odot}$~km\,s$^{-1}$\,yr$^{-1}$), indicating that the jet is powerful enough to drive the outflow. However, the outflow shows a two-component structure in our APEX and SMA observations. The SMA map of the CO emission at higher velocities (Figure~\ref{fig:int_sma}(d)) reveals molecular clumps lying exactly within a narrow cone being carved by the wiggling jet, providing strong evidence that the central collimated component is entrained by the jet. On the other hand, the wide-angle component has an opening angle reaching $30^{\circ}$, which could not be easily accounted for by the extremely collimated jet that is only gently wiggling within a few degree \citep{Marti93,Masque12}. Another model that could potentially produce a wide-angle outflow shell around a collimated jet is through jet bow-shocks which are created by sideway ejections from internal shocks within the jet \citep{Raga93,Masson93}.
The jet bow-shock model predicts distinct velocity structures in molecular outflows: extremely high velocity features \citep{Masson93}; the maximum velocities increasing with the distances from the central source in a position-velocity (PV) diagram cut along the jet axis \citep[i.e., the ``Hubble wedges'', see, e.g.,][]{Arce02}; a spur-like feature with the largest velocity dispersion at the largest distance to the central source in the PV diagram \citep{Masson93,Lee01}. Such velocity structures have been detected in both low-mass and high-mass outflows which are interpreted as jet bow-shock driven flows \citep[e.g.,][and references therein]{Qiu11}. The outflow velocities measured in the APEX and SMA maps are only a few km\,s$^{-1}$. We do not detect any high velocity ($v>10$~km\,s$^{-1}$) emissions in any of the CO lines. Figure~\ref{fig:pv} shows the APEX CO~(6--5) PV diagram cut along the jet/outflow axis. We cannot identify any PV structure that is predicted by a jet bow-shock model. Instead the PV pattern of the SW and redshifted lobe shows a concave structure which curves outward from the point of the central source position and the cloud velocity. Such a PV structure has been observed in wide-angle outflows in both low-mass and high-mass YSOs and is consistent with the scenario that the outflow is driven by a wide-angle wind \citep{Lee01,Qiu09a}. 

Therefore, the HH~80--81 outflow cannot be understood as a purely jet driven flow. We suggest that the central collimated component and the wide-angle component each traces part of the mass flow ejected from the high-mass YSO; the mass flow consists of a fast and collimated jet (previously detected in radio continuum) and a wide-angle wind (suggested by the wide-angle component of the molecular outflow and the wide-angle cavity seen in the \emph{Spitzer} image). It is worth noting that for low-mass outflows, the collimated component (or ``molecular jets'') are typically more than 10~km\,s$^{-1}$ faster than the wide-angle component, whereas in the HH~80--81 outflow, the collimated component is only $\sim$2~km\,s$^{-1}$ faster than the wide-angle component. Most recent ALMA observations suggest that low-mass outflows could be directly ejected from accretion disks, and thus the velocity difference between the two components manifests the difference in their launching radii on the disk \citep{Bjerkeli16,Lee17,Alves17,Tabone17,Guedel18}. Here for the HH~80--81 outflow, the velocity of the radio jet is of order 1000~km\,s$^{-1}$ \citep{Marti93,Marti95}, and as discussed above, the central collimated component of the outflow has a velocity of only $\lesssim$10~km\,s$^{-1}$ and should come from ambient material entrained or swept up by the jet. The velocity of the wide-angle wind is unknown, and could be estimated if extremely high angular resolution ($\lesssim$0.01\arcsec, or $\lesssim$17~AU), high sensitivity, and high fidelity observations capable of resolving the wind launching zone on the disk are available. Therefore the question about whether the wide-component of the outflow contains material directly ejected from the disk or entrained ambient gas, or both, remains open with the existing observations.

\subsection{Implications to a possible evolutionary picture for outflows in high-mass YSOs}
Low-mass outflows have been found to exhibit an evolutionary sequence in collimation: the outflow is highly collimated in the earliest Class 0 stages and continues to widen through late Class 0 to Class I and Class II stages \citep{Velusamy98,Arce06,Seale08,Velusamy14,Hsieh17}. The exact mechanism responsible for the outflow broadening is not fully understood \citep{Shang06,Offner11}. A possible explanation invokes a relatively slow wide-angle flow around a much faster and denser axial jet ejected from the star and disk system. At the earliest stages, only the jet component can puncture the infalling envelope; as the envelope loses mass through infall and outflow,  the wide-angle wind will break through and eventually become the main component; sideway splash of material from internal shocks may also contribute to broaden the outflow cavity\citep{Arce07,Frank14,Bally16}.

To account for the difference in morphology seen in some of the observed outflows in high-mass star-forming regions, \citet{Beuther05} suggested that the outflow opening angle continues to increase over time as the central high-mass YSO evolves from a protostellar stage to a hypercompact (HC) and ultracompact (UC) H{\scriptsize II} region, or alternatively, grows in mass equivalent to spectral types of mid/early-B to early-O types. It is still not well established that whether and how outflows in high-mass YSOs evolve \citep{Qiu08,Kuiper16}. The opening angle of the HH~80--81 outflow is $28^{\circ}$ measured from the \emph{Spitzer} image. The angle would be slightly larger if measured from the CO images (see Figures~\ref{fig:overview}--\ref{fig:CO65_chan}). This agrees with \emph{Spitzer} observations of outflow cavities in a sample of low-mass YSOs, and could be due to entrainment of material just beyond the wall into the cavity by the wide-angle wind \citep{Seale08}. Compared to some well shaped massive outflows with similar scales ($\sim$1~pc), the HH~80--81 outflow has a moderate opening angle and a dynamical age of $7\times10^4$~yr, which is larger than those of well collimated flows with dynamical ages $\lesssim10^4$~yr \citep{Beuther02b,Qiu09b,Zhang15}, and smaller than those of poorly collimated flows with dynamical ages of a few $10^4$--$10^5$~yr \citep{Shepherd98,Qiu09a}. This comparison seems to support an evolutionary sequence qualitatively similar to what is established for low-mass outflows, and further suggests that the mass ejection and accretion processes in the formation of early-B to late-O type stars could be similar to those in the formation of Sun-like stars. However, the well studied wide-angle outflows emanating from UC H{\scriptsize II} regions do not have an accompanying jet in existing observations \citep{Shepherd98,Qiu09a}. This is different from low-mass outflows, which are known to be associated with an axial jet from Class 0 to Class II stages \citep{Frank14,Bally16}. It is unclear why a jet component completely disappears in relatively later stages (e.g., the UC H{\scriptsize II} region stage) of high-mass star formation. The expansion of ionized gas and/or radiation feedback might play a role there \citep{Keto02,Kuiper16}.

\section{Summary} \label{sec:sum}
We have performed CO multi-line observations of the HH~80--81 outflow using the APEX, JCMT, and CSO, as well as high resolution CO and $^{13}$CO (2--1) observations using the SMA. We detect both wide-angle flows with an opening angle of about $30^{\circ}$, and clumps and knots following the path of the gently wiggling jet. Hence the HH~80--81 outflow is of the ``two-component'' nature, and the velocity structure suggests that each of the two components traces part of the mass loss process. The outflow mass and energetics estimated from the CO~(6--5) data are dominated by the wide-angle component, and are significantly lower than previous estimates based on low resolution CO~(1--0) and (2--1) observations, which were apparently affected by contamination from ambient cloud structures and other outflows in the region. Comparing the outflow with well shaped massive outflows available from the literature, we find that the opening angle of massive outflows continues to increase over dynamical ages of $10^3$ to $10^5$~yr. This is qualitatively similar to an evolutionary sequence established for low-mass outflows. However, there does exists difference in the sense that a jet component disappears in massive outflows at later stages, whereas low-mass outflows are always associated with an axial jet from Class 0 to Class II stages.

\acknowledgments K.Q. acknowledges supports from National Natural Science Foundation of China (Grant Nos. 11473011, U1731237, 11590781, and 11629302). K.Q. is supported by National Key R\&D Program of China No. 2017YFA0402600. This research used the facilities of the Canadian Astronomy Data Centre operated by the National Research Council of Canada with the support of the Canadian Space Agency.

\facility{Atacama Pathfinder EXperiment (APEX), James Clerk Maxwell Telescope (JCMT), Submillimeter Array (SMA), Caltech Submillimeter Observatory (CSO)}

\software{ORAC-DR pipeline, MIR (\url{https://github.com/qi-molecules/sma-mir}), MIRIAD \citep{Sault95}, RADEX code \citep{vdTak07}}

\appendix

\section{Comparison between the JCMT map and convolved APEX maps} \label{append:convolved}
\begin{figure}
\plotone{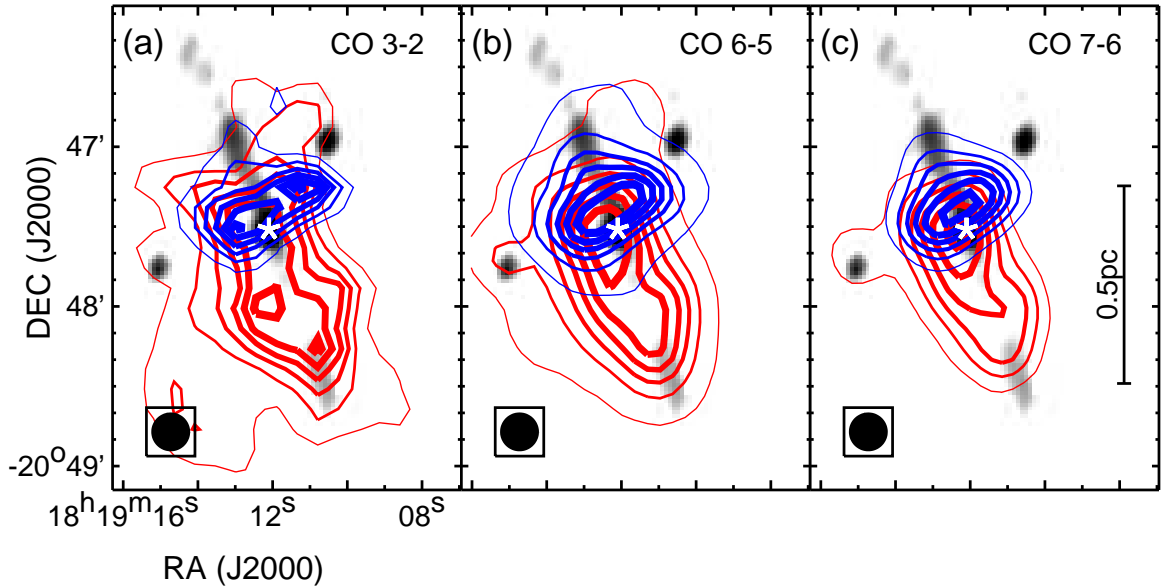}
\caption{Velocity integrated emissions in CO (3--2), (6--5), and (7--6). (a) The same as that for Figure \ref{fig:int_maps}(b). (b) The same as that for Figure \ref{fig:int_maps}(c), except that the map has been convolved to an angular resolution of 14.\arcsec5, and the peaks are 43.3 and 30.2~K\,km\,s$^{-1}$ for the blue and red lobes, respectively. (c) The same as that for Figure \ref{fig:int_maps}(d), except that the map has been convolved to an angular resolution of 14.\arcsec5, and the peaks are 30.5 and 20.4~K\,km\,s$^{-1}$ for the blue and red lobes, respectively. \label{fig:A}}
\end{figure}

Figure~\ref{fig:A} shows a comparison between the velocity integrated JCMT CO (3--2) map and the APEX CO (6--5) and (7--6) maps; the APEX maps have been convolved to the angular resolution of the JCMT map.

\section{CO (3--2) and (7--6) channel maps} \label{append:chan}
\begin{figure}
\plotone{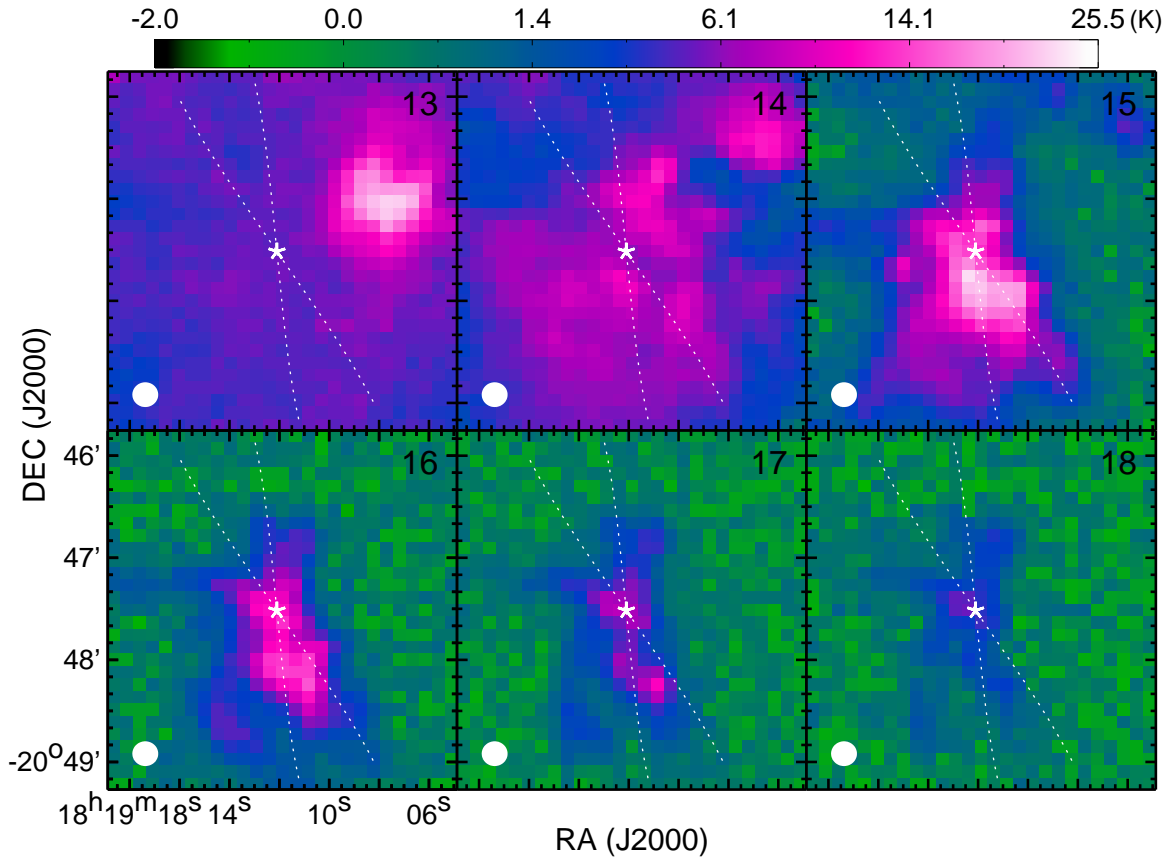}
\caption{Velocity channel maps of the CO (3--2) emission at 13 to 18 km\,s$^{-1}$. The pseudo color scale, as indicated by a color bar on top of the figure, visualizes intensities from $-2.0$ to 25.5~K in $T_{\rm A}^{\ast}$. Other symbols are the same as those in Figure \ref{fig:CO65_chan}. \label{fig:B1}}
\end{figure}

\begin{figure}
\plotone{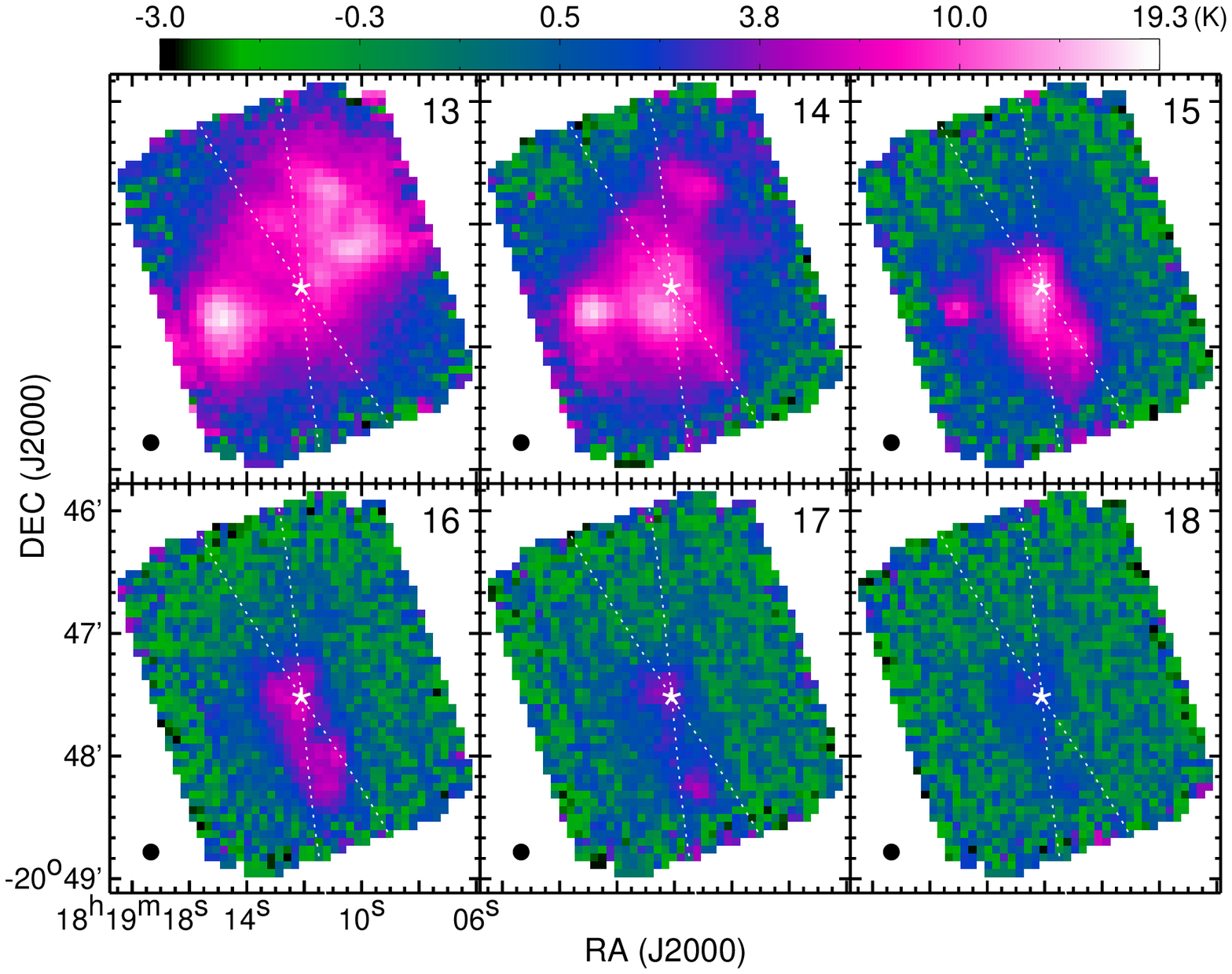}
\caption{Velocity channel maps of the CO (7--6) emission at 13 to 18 km\,s$^{-1}$. The pseudo color scale, as indicated by a color bar on top of the figure, visualizes intensities from $-3.0$ to 19.3~K in $T_{\rm A}^{\ast}$. Other symbols are the same as those in Figure \ref{fig:CO65_chan}. \label{fig:B2}}
\end{figure}

Figure~\ref{fig:B1} shows the velocity channel maps of the CO (3--2) emission, and Figure~\ref{fig:B2} shows the velocity channel maps of the CO (7--6) emission.  

\end{document}